\documentstyle[epsfig]{mn} 

\begin{document}

\title[Accretion stream mapping with genetically modified `fire-flies']
{Accretion stream mapping with genetically modified `fire-flies'}

\author[C. M. Bridge et al.]  
{C.M. Bridge$^1$, Pasi Hakala$^2$, Mark Cropper$^1$, Gavin Ramsay$^1$, \\ 
$^1$ Mullard Space Science Laboratory, University College London, Holmbury St. Mary, Dorking, Surrey, RH5 6NT\\
$^2$ Observatory, P.O. Box 14, Fin-00014, University of Helsinki, Finland}

\maketitle

\begin{abstract}
We apply an eclipse mapping technique using `genetically modified fire-flies'
to the eclipse light curves of HU~Aqr and EP~Dra. The technique makes as few
assumptions as possible about the location of accretion stream material,
allowing the emission to be located anywhere within the Roche lobe of the white
dwarf. We model two consecutive eclipses in the $UBVR_c$-band for HU~Aqr, and
four consecutive `white'-light eclipses for EP~Dra, to determine the changing
brightness distribution of stream material. We find fire-fly distributions which are consistent with accretion through a curtain of material in both HU Aqr and EP Dra, and show that the previously assumed two part ballistic and magnetic
trajectory is a good approximation for polars. Model fits to the colour band data of HU~Aqr indicate that the material confined to the magnetic field lines is brightest, and most of the emission originates from close to the white
dwarf. There is evidence for emission from close to a calculated ballistic
stream in both HU~Aqr and EP~Dra.We propose that a change in the stream density
causes a change in the location of the bright material in the accretion stream
in EP~Dra.
\end{abstract}

\begin{keywords}
accretion, -- binaries: eclipsing -- stars: individual: HU Aqr -- stars:
individual: EP Dra
\end{keywords}

\section{Introduction}

AM Hers, or polars, are a sub-class of the magnetic cataclysmic variable (CV)
interacting binaries. A main sequence secondary transfers material to a white
dwarf primary through an accretion stream, from mass overflow at the inner
Lagrangian point ($L_1$). The primary has a magnetic field of order
10$\--$200~MG, which prevents the formation of an accretion disk (as found in
other types of cataclysmic variables), and instead, at some threading radius
from the primary, the material is confined to follow the magnetic field lines
to accrete directly onto the white dwarf surface (see Cropper 1990 for a
review). A number of these systems have inclinations to the line of sight such
that the system is eclipsed by the secondary, and this can be used as a means
of isolating the emission from discrete parts of the system. In particular the
brightness distribution of material along the accretion stream can be inferred
as successive sections of the accretion stream material are eclipsed.

Attempts to determine the brightness distribution of the accretion stream
material has evolved in complexity from initially one-dimensional streams
confined to the orbital plane (Hakala 1995), to three dimensional tubes
carrying material far out of the orbital plane (Kube, G\"ansicke \& Beuermann
2000; Vrielmann \& Schwope 2001). These methods have been applied to both
emission line (Kube, G\"ansicke \& Beuermann 2000; Vrielmann \& Schwope 2001)
and continuum observations (Hakala 1995; Harrop-Allin et al. 1999b, 2001;
Bridge et al. 2002). The assumption common to all previous model techniques is
that of a stream trajectory determined prior to the modelling process, and
fixed for the duration of the model. In an attempt to remove as many of the
assumptions about the stream location as possible, a technique has been
developed by Hakala, Cropper \& Ramsay (2002), which makes fewer assumptions
about the location of bright stream material. In principle, stream material can
be located anywhere within the Roche lobe of the primary.

The application of the technique to synthetic data sets was demonstrated in
Hakala et al. (2002), and here we apply the method to optical light curves of
two eclipsing polars: EP~Dra and HU~Aqr. The two systems show variations in the
brightness and trajectory of the accretion stream over the timescale of the
orbital period (Bridge et al. 2002, 2003), with EP~Dra also showing a variation
in brightness over a longer phase range. This variation appears to be related
to the brightness of the accretion stream, and is attributed to a combination
of cyclotron beaming and absorption in an extended accretion curtain (Bridge et
al. 2002, 2003).

The eclipse mapping technique based on the model of Harrop-Allin et al. (1999a)
was found (by Bridge et al. 2002) to be particularly sensitive to the input
parameters and the signal-to-noise ratio of the data used, when applied to the
selected light curves of HU~Aqr, and hence restricted the interpretation of the
results. We therefore apply this new model technique to the EP~Dra and HU~Aqr
light curves in an attempt to circumvent these limitations.

\section{Data sets}

\begin{table}
\caption{Summary of the observations of EP~Dra and HU~Aqr. Cycle numbers are
with respect to the ephemeris of Schwope \& Mengel (1997) for EP~Dra and
Schwope et al. (2001) for HU~Aqr.}
\begin{center}
\begin{tabular}{lclrr}
\hline
& Cycle	& Eclipse centre& Observation	\\
& number& (TDB)		& length (s)	\\
&	& (2450000.0+)	&		\\
\hline	
EP Dra& 56962	& 1820.37575	& 1496 	\\
& 56976 & 1821.39294	& 2400 	\\
& 56977 & 1821.46559	& 2111 	\\
& 56978 & 1821.53825    & 1393  \\
HU Aqr& 29994 & 1821.43744    & 2400  \\
& 29995	& 1821.52426	& 1216 	\\
\hline
\end{tabular}
\end{center}
\label{tab:obs}
\end{table}

Light curves of EP~Dra and HU~Aqr were obtained on the nights of 2000 October
2/3 and 2000 October 3/4. The observations were taken using the superconducting
tunnel junction instrument S-Cam~2 (see Perryman et al. 2001 and reference
therein). S-Cam~2 provides spectral information through the ability to record
the energy of each incident photon, as well as the time of arrival and position
on the detector array. The HU~Aqr light curves presented in Bridge et
al. (2002) have now been split into energy ranges that more closely resemble
those of the Johnson-Cousins {\it UBVR$_c$}-bands (Bessell 1990, and references
therein), as for EP~Dra and detailed in Bridge et al. (2003). The $U$-band
represents the range 340$\--$400\,nm, the $B$-band is 390$\--$490\,nm, the
$V$-band is 500$\--$600\,nm and the $R_c$-band is
590$\--$680\,nm. Table~\ref{tab:obs} gives the cycle number, start time (TDB)
and total observation length for those light curves used in the modelling
presented here. EP~Dra was too faint to provide high enough signal-to-noise
ratio light curves when divided into $UBVR_c$-band energy ranges (V$\approx$17
for EP~Dra compared to V$\approx$15 for HU~Aqr), therefore we use the
`white'-light curves (340$\--$680\,nm).

The effect of low signal-to-noise ratio data on the fits is that the fire-fly
swarm is broadened. In light of these considerations, we have chosen cycles
29994 and 29995 of HU~Aqr as having the best signal-to-noise ratio and
sufficient phase range, and cycles 56962, 56976, 56977 and 56878 from
EP~Dra. The choice of consecutive orbits of light curves means that we can
compare the changing brightness and eclipse profile of the accretion stream in
consecutive eclipses. The observed light curves were binned into 4\,s time
bins. This is to reduce the computational time required for the model. The
EP~Dra light curves have been truncated compared to those presented in Bridge
et al. (2003), to cover the eclipse phases only, and hence the `trough' feature
identified in the longer phase coverage light curves is not evident.

\section{The Model}

\subsection{Overview}

The model is described in detail in Hakala et al. (2002), so we provide only a
brief description here. The model uses a number of bright emission points,
dubbed `fire-flies', that are free to move within the Roche lobe of the
primary. Each fire-fly has an angle-dependent emission, given by $F_{fly} =
F_{0}+A\mathrm{cos(}\alpha\mathrm{)}$, where $F_{0}$ is the minimum brightness
of a fire-fly, $A$ is the amplitude of angular dependence and $\alpha$ is the
angle between the primary, the fire-fly and the observer. This angular
dependence was included for two reasons: firstly to account for the effect of
X-ray heating of the side of the accretion stream facing the primary, and
secondly to mimic the effect of the optical thickness of the stream near the
eclipse phases. The model evolves a best fit light curve and hence fire-fly
distribution, by summing the brightness of the fire-flies visible at each phase
of the light curve.

The model creates a number of initially random fire-fly `swarms', and each of
these swarms is used to create a model light curve. The swarms are then evolved
towards a final solution using a genetic algorithm (GA; see Charbonneau 1995
for a review) which evaluates the goodness of fit of each model light curve,
and uses the swarms of fire-flies corresponding to the best fitting light
curves to create the next generation of fire-fly swarms. This process is
repeated until the model converges, or a pre-set number of generations has
passed.

The location of the fire-flies defines an emission volume, and so the precise
location of an individual fire-fly within that volume is not necessarily
important or unique. The bright sections of stream are seen as a larger number
of fire-flies concentrated in a smaller volume.

\subsection{Evaluation of the fit}

The goodness-of-fit of a particular model light curve is a combination of the
$\chi^2$-fit of the fire-fly generated model light curve to the observed light
curve, plus an optional regularisation term derived from a self-organising map
(SOM; Kohonen 1990). The SOM places a curve through a given swarm, with ends
located near the primary and $L_1$. The fire-flies are constrained to prefer to
lie at minimal distance from this curve, and hence those swarms that follow the
shape of the curve more closely have correspondingly better fitness
functions. The details of the application of a SOM to the fire-fly model are
given in Hakala et al. (2002). They did not specifically explore the effects of
two important parameters in the regularisation, that of the neighbourhood
kernel width and the number of nodes (or sections) in the curve. As we now
apply this technique to `real' data for the first time, we briefly explore the
effects of varying these two parameters.

\begin{figure}
\centerline{\epsfig{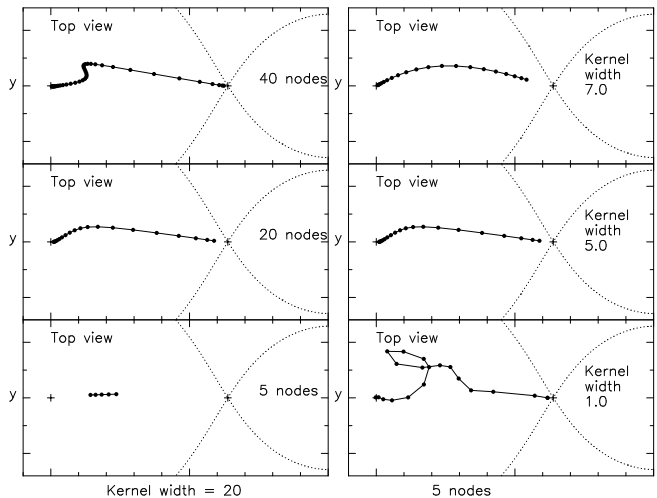}}
\caption{The left hand column shows regularisation curves for different numbers
of nodes and the right hand column different neighbourhood kernel widths. The
middle two curves represent the values used for the modelling presented here.}
\label{fig:huaqrnodes}
\end{figure}

For a test dataset, we have chosen the `white'-light curve of HU~Aqr cycle
29995 (Bridge et al. 2002), and apply the same input parameters to the model,
varying either the kernel width or the number of
nodes. Fig.~\ref{fig:huaqrnodes} shows the resulting regularisation curves. To
create a best-fit curve through a particular fire-fly swarm, the model selects
a random fire-fly and finds the nearest node, this node is then moved towards
the fire-fly. Each adjacent node in the curve is also moved; however the
amplitude of the movement is reduced for each subsequent node. The amount by
which the movement is reduced is set by the kernel width. For each swarm this
process of moving the nodes is repeated for around 10$\--$20 times the number
of fire-flies in the swarm. The use of a large kernel width will propagate
larger amplitude movements further along the curve, leading to a smoother, or
`stiffer', curve. The number of nodes has a similar effect: a large number of
nodes leads to localised turns in the curve as the amplitude of the node
movement decreases in a relatively short distance along the stream. In general,
the ratio of the number of nodes to the kernel width affects the strength of
the regularisation by altering the stiffness of the regularisation curve.

The regularisation curve is a constraint on the fire-fly distribution, and the
form of the curve is important as it can impose preconceived ideas of a stream
trajectory on to the fire-fly swarms. The choice of the most appropriate values
for the kernel width and the number of nodes is set by the need to allow the
model to create a stream that is sufficiently stiff as to reproduce a
`physically realistic' accretion stream, but not too stiff as to overly
constrain the evolution of the fire-fly distributions. The effect of the
regularisation can be significant in shaping the fire-fly swarm, as can be seen
from Fig.~\ref{fig:huaqrnodes}. This will influence the model results and their
interpretation. However, we can make progress towards an optimal means of
regularisation by excluding the more extreme regularisation parameters, such as
that found for small kernel widths or large numbers of nodes. With this in
mind, we have chosen to use 20 for the number of nodes and 5.0 for the
neighbourhood kernel width, as in the middle plots of
Fig.~\ref{fig:huaqrnodes}. These are the same as those used by Hakala et
al. (2002) for their fits to synthetic data. When interpreting the results in
Section~\ref{sec:results} the choice of kernel width and node number, and the
possible influence on the interpretation of the resulting fire-fly
distributions, must be borne in mind.

\subsection{Fixed parameters}
\label{sec:parameters}

A number of fixed parameters are used in the model. These include parameters
describing the binary system, such as the mass ratio and binary inclination,
and parameters such as the number of swarms, number of fire-flies per swarm and
the brightness dependence of the fire-flies. The number of fire-flies and
swarms chosen must reflect the need to introduce enough diversity into the
population to explore the whole parameter space and reach a unique solution,
while the number of iterations must be sufficient to reach convergence. In
practice, the number of iterations is also constrained by the computational
time required for a model fit.

\begin{figure}
\centerline{\epsfig{file=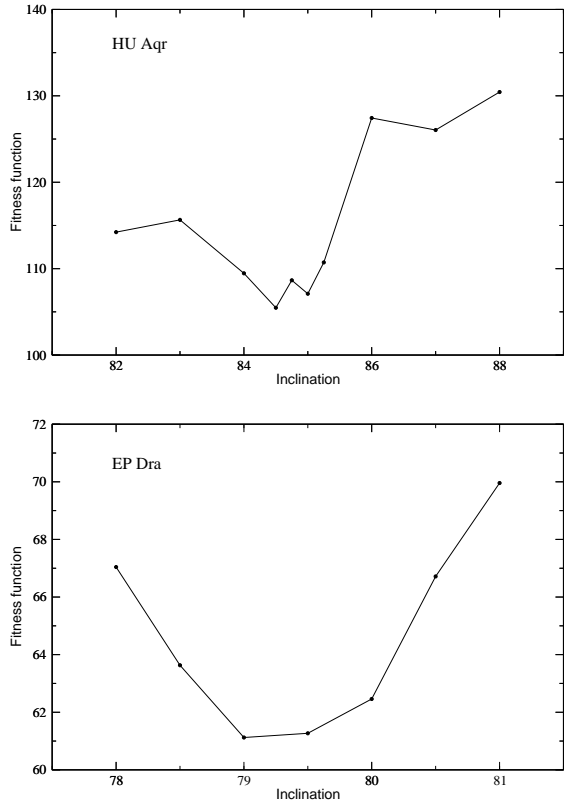,width=8.0cm,angle=0}}
\caption{The upper plot shows the fitness function for a range of values of
inclination for the $V$-band of cycle 29995 of HU~Aqr, and the bottom plot for
cycle 56976 of EP~Dra.}
\label{fig:constrain}
\end{figure}

Several of the parameters can potentially be determined using the model fitting
itself. The inclination $i$ and the brightness dependence of the fire-flies,
which affects $F_{fly}$, can be varied to determine the best value from the
fitness function. This was demonstrated in Hakala et al. (2002) for synthetic
data sets, and here we apply it to observed light curves (see
Section~\ref{sec:constrain}). The brightness dependence is varied through the
user input value of the emissivity ratio $\varepsilon = 1/(F_0-A)$. This
defines the brightness of the directly opposing faces of the fire-fly --
i.e. the brightness of the fire-fly in the direction facing the primary,
compared to that facing directly away from the primary. A third parameter is
used to mimic the effects of the accretion region on the primary: the value for
the brightness of the accretion region is fixed at the ingress, and then, using
a `pole trend' parameter, the angular dependence of the emission from this
region is accounted for by increasing this value linearly with phase to match
the egress brightness. In those observations covering a small phase range this
was found to have little effect, as the model effectively uses fire-flies to
compensate for an inappropriate brightness at the accretion region. However it
is important for HU~Aqr cycle 29994 which has a longer egress phase range (see
Section~\ref{sec:huaqrresults}). The pole trend cannot, of course, account for
flaring seen in the longer egress light curves.

\section{Results}
\label{sec:results}

\begin{figure*}
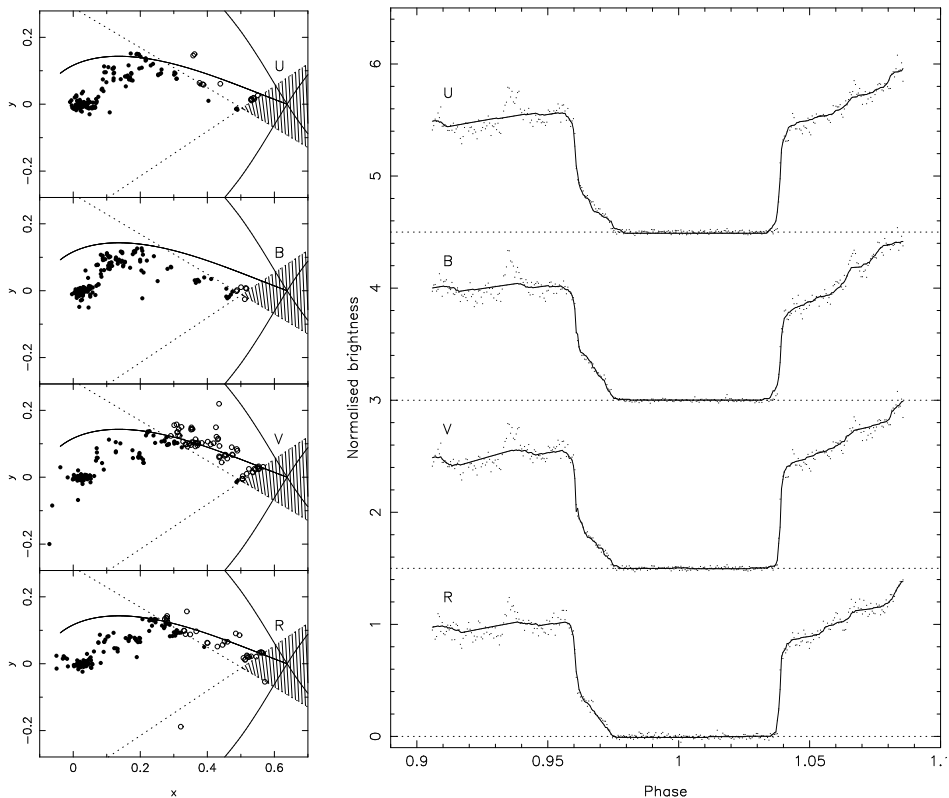

\epsfig{file=./huaqr_03_flies.ps,width=4.0cm,angle=0}
\hspace{3.0mm}
\epsfig{file=./huaqr_03_fits.ps,width=8.0cm,angle=0}
\caption{The left hand panel shows the fire-fly distributions for the model
fits to the light curves of HU~Aqr for cycle 29994, looking down onto the
orbital plane. The Roche lobe and a ballistic accretion stream are represented
with solid lines, with two dotted lines to indicate those parts of the system
observed in the phase range of the light curves. The coordinates are centred on
the white dwarf at 0,0. The right hand panel shows the model fits (solid line)
to the observed {\it $UBVR_c$}-band light curves (dots) of HU~Aqr. Each light
curve and fit is offset vertically by 1.5 for clarity.}
\label{fig:huaqr03}
\end{figure*}

\begin{figure*}
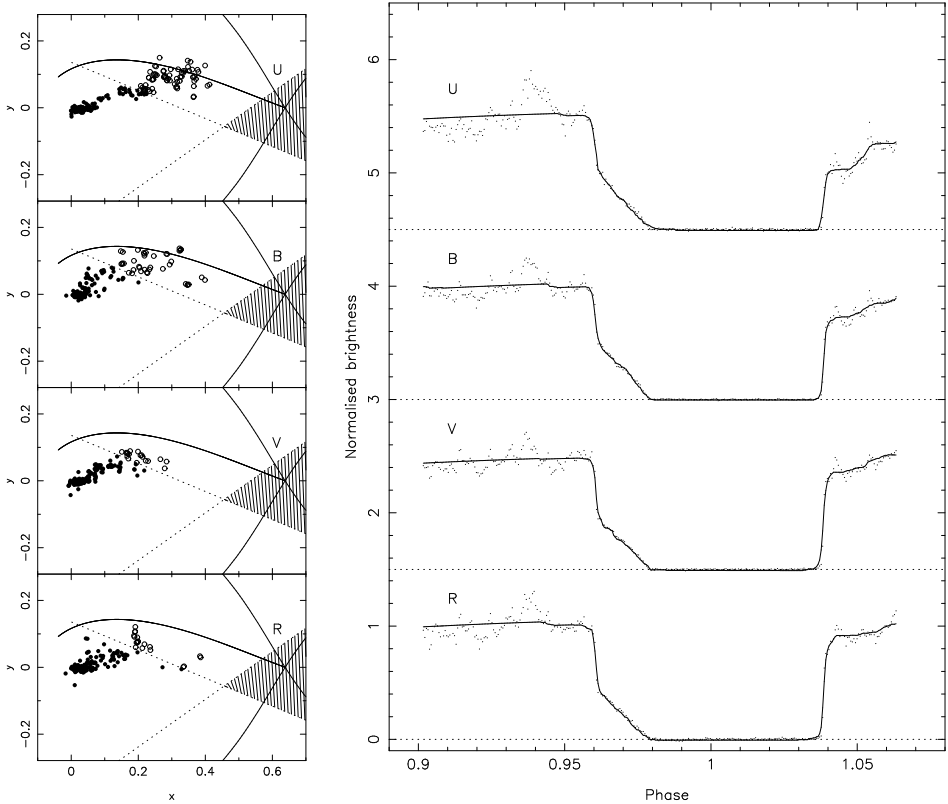

\epsfig{file=./huaqr_04_flies.ps,width=4.0cm,angle=0}
\hspace{3.0mm}
\epsfig{file=./huaqr_04_fits.ps,width=8.0cm,angle=0}
\caption{As for Fig.~\ref{fig:huaqr03}, but for HU Aqr cycle 29995.}
\label{fig:huaqr04}
\end{figure*}

\begin{figure*}
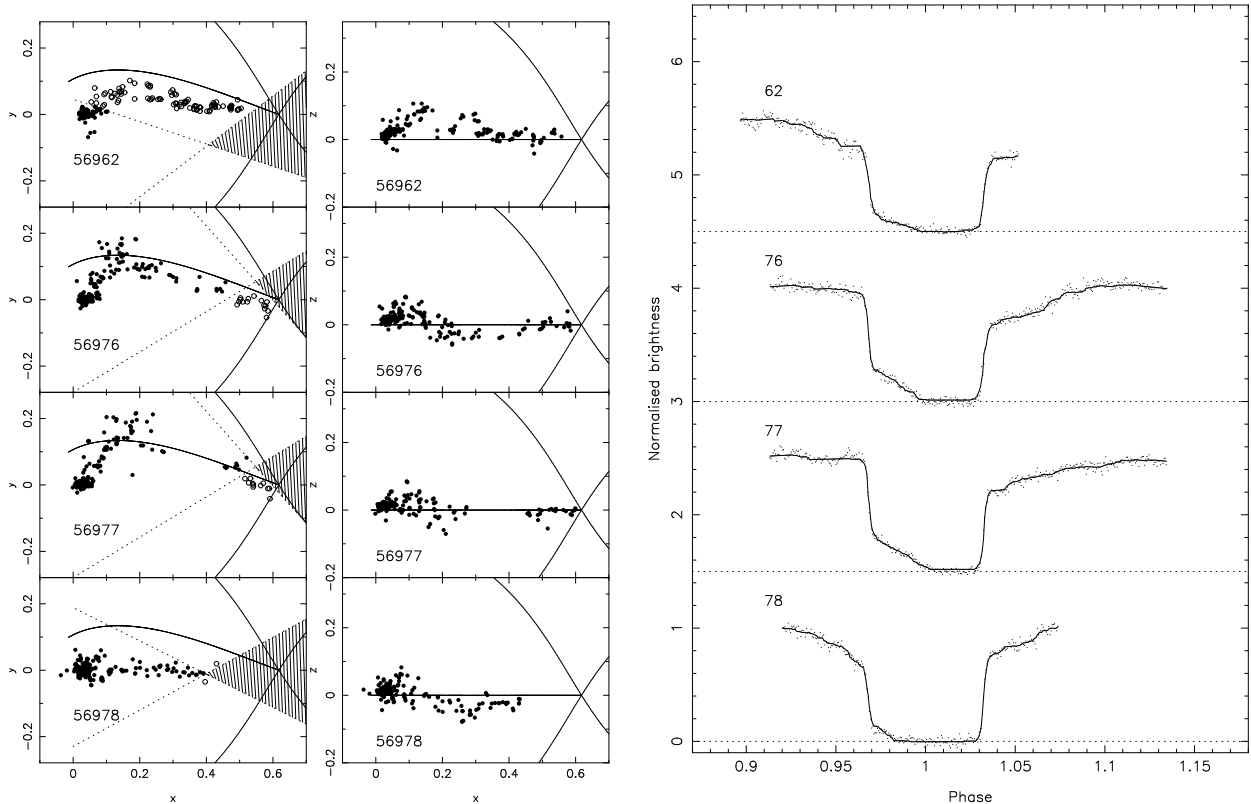

\epsfig{file=./epdraflies.ps,width=8.0cm,angle=0}
\hspace{3.0mm}
\epsfig{file=./epdrafits.ps,width=8.0cm,angle=0}
\caption{Left two panels: the fire-fly distributions for the model fits to the
`white'-light curves of EP~Dra. The first panel shows the view looking down
onto the orbital plane, and the second panel the view parallel to the orbital
plane. The Roche lobe and a ballistic accretion stream are represented with
solid lines, and the first panel includes two dotted lines to indicate
those parts of the system observed in the phase range of the light
curves. Right hand panel: the model fits (solid line) to the different cycles of EP~Dra. Each light curve and fit is offset vertically by 1.5 units for clarity.}
\label{fig:epdraresults}
\end{figure*}

\subsection{Constraining system parameters}
\label{sec:constrain}

We can constrain the binary inclination using the fire-fly model
fits. Fig.~\ref{fig:constrain} shows the fitness function for a range of values
of $i$ for EP~Dra cycle 56976 and the $V$-band of HU~Aqr cycle
29995. Fig.~\ref{fig:constrain} indicates a value of $i=79\--80^{\circ}$ for
EP~Dra and we use $i=79.5^{\circ}$ for subsequent modelling. This value is in
good agreement with the value of $i=80^{\circ}$ derived by Remillard et
al. (1991) from their observed eclipse duration. For HU~Aqr we used a value of
$i=85^{\circ}$ which is in good agreement with the value from Schwope et
al. (2001) of $85.6^{\circ}$.

Constraints can be placed on the values of $\varepsilon$ also, and we find here
that the values are constrained within approximately $\pm 0.1$ of the optimal
value. The value of $\varepsilon$ is also linked to the value of the pole
trend. A decrease in the value of $\varepsilon$ means that the fire-flies are
brighter on the side facing the white dwarf, seen after eclipse on accretion
stream egress. This is also when the model is providing the most contribution
to the cyclotron emission through the pole trend parameter. The phase
dependence of the emission from the fire-flies therefore implies that this
parameter $\varepsilon$ will most likely play a more important role in long
phase range light curves, as suggested by Hakala et al. (2002).

\subsection{HU Aqr}
\label{sec:huaqrresults}

The results of the fire-fly modelling for cycles 29994 and 29995 of HU~Aqr are
shown in Figs.~\ref{fig:huaqr03} and \ref{fig:huaqr04} respectively. The
left-hand panel shows the distribution of fire-flies, and the right-hand panel
the corresponding model fits to the observed light curves. The white dwarf is
located at the origin. A large concentration of fire-flies indicates a bright
region. The dotted straight lines in the left-most panels of
Figs.~\ref{fig:huaqr03} and \ref{fig:huaqr04} define the regions of the system
seen from the first to last phases in the light curves. The shaded area is
therefore never seen, and thus any fire-flies in this region make no
contribution to the model fits and are excluded from the plot. Those fire-flies
that are seen on both ingress and egress are shown as filled circles, while
those seen either on ingress or egress alone are represented as hollow
circles. The high value of $i=85.0^{\circ}$ means that the fit is constrained
mainly in the orbital (x-y) plane and less so in the x-z plane. We do not
include a plot of the x-z plane fire-fly distribution for this reason.

The phase range of the observed light curve for cycle 29994 is longer on egress
than that for cycle 29995, which means that the pole trend is more important
for this cycle.  The $B$-band fit of cycle 29995 does require a higher value of
the pole trend than the $UVR$-band fits. The shorter egress in cycle 29995
means that changing the value of $\varepsilon$ has no effect, and the same
value of $\varepsilon=0.2$ is therefore used for all four bands. (This means
that the fire-flies are 5 times brighter facing the primary, than when facing
away.)  For cycle 29994 the value is different in the different bands because
of the greater influence of the cyclotron emission on the model fits after the
egress of the accretion region. The values for the $UBVR$-bands were 0.25, 0.2,
0.3 and 0.2 respectively.

In Fig.~\ref{fig:huaqr03} there is a concentration of fire-flies close to the
white dwarf, and there are also a number of fire-flies defining the trajectory
of material from $L_1$ to the white dwarf. However, most of the fire-flies are
located in the region where we expect the material to be channeled by the magnetic
field lines of the white dwarf. The increased number of fire-flies along the
indicated ballistic trajectory in the $V$-band and $R$-band compared to the
$U$-band and the $B$-band appears to be a result of the increase in the model
light curve around $\phi\approx 0.937$. This increase coincides with the large
increase in the observed light curve centred on $\phi\approx
0.935$. Fig~\ref{fig:huaqr04} indicates that for cycle 29995, most of the
emission comes from the region close to the primary and nearly all the
fire-flies are located where we expect the material to be threaded onto the
magnetic field lines of the primary.

The differences between the bands for HU~Aqr cycle 29994 are noticeable in the
ballistic stream. There are more fire-flies closer to $L_1$ in the $V$-band and
$R$-band, and more in the region where we expect material to be coupled to the
magnetic field lines in the $U$-band and $B$-band. There is little emission
from a ballistic trajectory in all the bands of cycle 29995, and there are few
$V$-band and $R$-band fire-flies close to the calculated ballistic trajectory
in cycle 29995. Instead the fire-flies are located in the region where we
expect material to be coupled to the magnetic field lines. Further to this, in
the case of the $V$-band and $R$-band, the fire-flies are even closer to the
primary and would be expected to have been lifted further out of the orbital
plane along the magnetic field lines.

A lack of material near the indicated ballistic trajectory in cycle 29995 is
either because there is a lack of material actually located in this region, or
that it is significantly fainter compared to the material located near the
white dwarf. In this case the model will preferentially place the fire-flies in
the brighter region. The fact that a constant $\varepsilon$ is used for the
different bands means also, that any small variations in the cyclotron emission
will be compensated for by the model using fire-flies at the white dwarf to
mimic the smaller variations.

Model fits to synthetic data created from a given fire-fly distribution have
been shown in Hakala et al. (2002) to reproduce the original breadth of the
accretion material well. We therefore expect the broad accretion stream
indicated in the fire-fly distributions to reflect the true location of stream
material. That this broad width of material is located where we expect it to be
confined to the magnetic field lines implies that the material is threaded on
to many different field lines towards the calculated ballistic trajectory
(indicated by a solid line in the figures). The distributions in the different
bands are similar for both cycles; however there are more fire-flies defining
the early sections of the stream trajectory in cycle 29994. The fire-flies in
the $U$-band of cycle 29995 trace a swarm that appears to originate from the calculated ballistic trajectory.

\subsection{EP Dra}

The fits for EP~Dra are shown in Fig.~\ref{fig:epdraresults}. We show four
eclipses, one from the first night and three consecutive eclipses from the
second night. The fire-fly distributions in Fig.~\ref{fig:epdraresults} show
emission located close to a calculated ballistic trajectory, in cycles 56962
and 56976 with the consequence that there is a larger relative decrease in the
light curve prior to the accretion region and white dwarf ingress than cycles
56976 and 56977. This is seen in the fire-fly distributions as a larger number
of fire-flies located close to the calculated ballistic trajectory. In cycle
56978 the rapid decrease in flux prior to the rapid eclipse of the accretion
region and white dwarf indicates that material is located towards the ends of
the trajectory, that is close to the white dwarf and $L_1$. As the model
reaches total eclipse earlier in phase compared to the preceding two cycles,
there is evidently no material located far from the line-of-centres towards the
calculated ballistic trajectory, the model places the fire-flies along the line
joining $L_1$ to the white dwarf. The possibility exists that the accretion
stream material is threaded at $L_1$ and the cycle 56978 fire-fly distribution
indicates that this may be the case, however it is considered unlikely.

The fire-fly distribution in cycle 56978 is expected to be very different from
that of the preceding two cycles as the light curves are very different. The
fire-flies indicate that the brightest region is close to the white dwarf, with
material along the line-of-centres. This is seen in the light curves as a
gradual decrease in emission prior to the eclipse of the accretion region, and
a relatively faint stream after this eclipse, when compared to the previous
cycles. The increased brightness near the white dwarf in this cycle may also be
compensation for a larger difference in the brightness between the ingress and
egress levels of the light curve, due to a change in the cyclotron emission.

The distribution of fire-flies for cycles 56976 and 56977 show that there is
material located in a relatively large region towards the end of the calculated
ballistic accretion stream. This region is located closer to the primary than
for the HU~Aqr fire-flies, which is indicative of either a weaker magnetic
field or a higher mass transfer rate. A weaker magnetic field explanation is
consistent with the derived field strengths for HU~Aqr of $B=37$~MG (Glenn et
al. 1994) and EP~Dra of $B\approx16$~MG (Schwope \& Mengel 1997).

The distribution of fire-flies in the z-direction is better constrained for
EP~Dra than HU~Aqr, because of the lower inclination of
$i=79.5^{\circ}$. However, the appearance of fire-flies below the orbital plane
may be indicative of a lack of constraint in the z-direction. The extent of the
fly distribution in the z-direction towards the white dwarf in all cycles
possibly indicates accretion at locations nearer the equator on the primary
(larger colatitude $\beta$) as the material does not have far to travel along
the magnetic field lines.

\section{Discussion}

Our results show that the previously assumed two part ballistic plus magnetic trajectory is a good approximation to the flows delineated by the fire-fly distributions.

\subsection{Changing stream brightness}

Together, the three EP~Dra eclipses from the second night (56976, 56977 and
56978) show a change in the location of bright material, which is related to a
change in the observed eclipse profile, and thus a change in the
brightness of the accretion stream material. This change is not necessarily
directly indicative of a change in the temperature of this material as the
brightness of stream material is directly dependent upon both the density and
temperature structure of the stream.

The change in brightness of the ballistic section of the accretion stream
between cycle 56976 and 56977 can be explained if material is stripped from the
ballistic stream in many places along the ballistic trajectory. This material
is channeled onto the magnetic field lines of the white dwarf, resulting in a
decrease in the amount of material in, and hence brightness of, the ballistic
stream. We do not necessarily see this material defined by the fire-flies
because it is fainter than the region where most of the material is coupled to
the magnetic field lines. As this coupled material reaches the white dwarf
there is an increase in the emission from this region of the system, as in
cycle 56978. Alternatively, the change in brightness may be caused by a change
in the rate of loss of material from the secondary resulting in a change in the
amount of material along the ballistic section of the stream. The time-scale
for material to travel from $L_1$ to the white dwarf is $\approx$50~minutes,
half the orbital period. Therefore, changes in the amount of material along the
accretion stream are possible between consecutive cycles.

\subsection{Stream heating}

The distribution of fire-flies in the different bands for HU~Aqr is indicative
of stream heating processes in both a coupling region and near the white
dwarf. This was found in Bridge et al. (2002) for HU~Aqr, and is predicted by
the theoretical models of Ferrario \& Wehrse (1999). The location of the bluer
fire-flies further from the primary in the $U$-band of cycle 29995 is
indicative of irradiated stream material. The accretion stream will be strongly
irradiated by X-ray emission as it reaches the threading region and rises from
the orbital plane along the magnetic field lines. This material is located
further away from the white dwarf by the model, where the observed emission is
brightest. Alternatively, the bright $U$-band emission may indicate that the
material is heated by processes in the magnetic coupling region.

\subsection{Comparison with previous results}

The HU~Aqr fire-fly distributions are consistent with the location of the
bright regions in the modelling of Bridge et al. (2002) and Harrop-Allin
(1999), which show variations in the brightness of the threading region and
regions near the white dwarf. The one-pole accretion models of Harrop-Allin
(1999) show that the magnetically confined regions of the accretion stream are
the brightest, with the $U$-band and $B$-band emission being greatest towards
the threading region. In particular the brightness distribution resembles the
fire-fly distributions of cycle 29994 with emission from the ballistic part of
the stream.

The threading region has been observed to move significantly between cycles
(Glenn et al. 1994; Bridge et al. 2002). The fire-flies will not necessarily
indicate a movement unless it is significantly larger than the width of the
fire-fly swarm. Bridge et al. (2002) found a change in the threading radius
between the two cycles modelled there, however the HU~Aqr cycle 29994 modelled
here was not included and in fact represents the intermediate cycle. Figure 5
of Bridge et al. (2002) shows the superimposed eclipse profiles of the HU~Aqr
cycles observed on the same nights. The figure indicates that the two cycles
29994 and 29995 have different eclipse profiles and hence different threading
radii.

Bridge et al. (2003) suggested the presence of an extended curtain of material
as being the cause of absorption over an extended phase range, and being at
least partially responsible for the trough feature seen in the light
curves. The fire-fly distribution in Figure~\ref{fig:epdraresults} supports the
idea that accretion material is threaded by many field lines. Schwope \& Mengel
(1997) suggested the presence of an accretion curtain, and their evidence for
an extended accretion arc on the surface of the white dwarf further supports
the idea that material is threaded onto many field lines.

\section{Conclusions}

We have applied the technique of eclipse mapping using genetically modified
fire-flies to the eclipse lights curves of EP~Dra and HU~Aqr. The modelling
shows that the technique is applicable to relatively good signal-to-noise ratio
light curves of adequate phase range. These distributions of fire-flies show
that the previously assumed ballistic free-fall plus magnetically confined
streams are a good approximation for the accretion streams in polars.

We applied the model to $UBVR_c$-band light curves of HU~Aqr cycles 29994 and
29995, and `white'-light curves of EP~Dra cycle 56962, 56976, 56977 and
56978. We have demonstrated that the technique will distinguish regions of
brightness in the different colour bands. This may indicate different
temperatures or densities, and hence heating and/or cooling processes in the
accretion stream or the location of photoionised material. 

The fire-flies distributions in both HU~Aqr and EP~Dra show the accretion
stream to be brightest near to the white dwarf. A possible threading region is
present at which the fire-fly distributions appear to deviate from a calculated
purely ballistic trajectory, and this region is seen to be broad, implying
accretion along many field lines.

Differences in the location and concentration of fire-flies between the three
cycles, 56976, 56977 and 56978, of EP~Dra indicate a changing brightness
distribution between the cycles. This could be the result of a change in the
temperature of the accretion stream material, or a change in the amount of
material stripped from the ballistic trajectory and coupled to the magnetic
field lines of the primary.

\section*{ACKNOWLEDGMENTS}

We wish to thank members of the Research and Scientific Support Department of
the European Space Agency at ESTEC for the use of the S-Cam~2 instrument.

\end{document}